\documentclass[conference, 10pt]{IEEEtran}

\usepackage{my_style}
\makeatletter
\def\ps@headings{
\let\@oddhead\@empty
\let\@evenhead\@empty
\def\@oddfoot{\@IEEEheaderstyle\hfil}%
\def\@evenfoot{\@IEEEheaderstyle\thepage\hfil\hbox{}}
}
\def\ps@IEEEtitlepagestyle{
\let\@oddhead\@empty
\let\@evenhead\@empty
\let\@evenfoot\@empty
}
\makeatother

\begin{document}
\setlength{\textfloatsep}{5pt}

\title{AttCDCNet: Attention-enhanced Chest Disease Classification using X-Ray Images}

\author{Omar H. Khater, Abdullahi S. Shuaib, Sami Ul Haq, Abdul Jabbar Siddiqui\\
\thanks{O. H. Khater,  A. S. Shuaib, and S. Ul Haq are with the Computer Engineering Department. A. Siddiqui is with the Computer Engineering Department and SDAIA-KFUPM Joint Research Center for Artificial Intelligence, KFUPM}
}

\maketitle

\begin{abstract}
Chest X-rays (X-ray images) have been proven to be effective for the diagnosis of chest diseases, including Pneumonia, Lung Opacity, and COVID-19. However, relying on traditional medical methods for diagnosis from X-ray images is prone to delays and inaccuracies because the medical personnel who evaluate the X-ray images may have preconceived biases. For this reason, researchers have proposed the use of deep learning-based techniques to facilitate the diagnosis process. The preeminent method is the use of sophisticated Convolutional Neural Networks (CNNs). In this paper, we propose a novel detection model named \textbf{AttCDCNet} for the task of X-ray image diagnosis, enhancing the popular DenseNet121 model by adding an attention block to help the model focus on the most relevant regions, using focal loss as a loss function to overcome the imbalance of the dataset problem, and utilizing depth-wise convolution to reduce the parameters to make the model lighter. Through extensive experimental evaluations, the proposed model demonstrates exceptional performance, showing better results than the original DenseNet121. The proposed model achieved an accuracy, precision and recall of \textbf{94.94\%}, \textbf{95.14\%} and \textbf{94.53\%}, respectively, on the COVID-19 Radiography Dataset.
\end{abstract}


\section{Introduction}
Respiratory diseases are a very serious public health problem that many people in the world suffer from. The diagnosis of such diseases is very important for proper patient care and therapy. X-ray images have become widely used as an easy and cost-effective way to promptly identify the unique patterns associated with respiratory diseases, enabling their diagnosis. Despite this, the precise and automatic diagnosis of such diseases from X-ray images appears to be a challenge because of the complexity and fluctuating nature of the disease manifestation \cite{arulananth2024classification}.

Over the past few years, various medical image analysis tasks, such as predicting Pneumonia from X-ray images, have demonstrated promising outcomes, especially through the use of Convolutional Neural Networks (CNNs)\cite{pereira2020covid}\cite{santosh2020truncated}. CNNs possess the capacity to acquire detailed information from images, enabling them to identify patterns that indicate the presence of respiratory diseases. DenseNet121\cite{huang2017densely}, a popular  CNN architecture, has shown excellent performance in medical image processing tasks thanks to its dense connections and efficient feature propagation \cite{arulananth2024classification}.

In this sense, the proposed work aims to determine the feasibility of early automated classifications of COVID-19, viral pneumonia, lung opacity, and normal cases based on X-ray images. The proposed model (AttCDCNet) builds upon and enhances the DenseNet121 architecture in order to detect whether the patient is healthy or has a chest disease. The novelty lies in introducing several enhancements designed to upgrade the performance and efficiency of the stated sota model. The proposed network has been capable of performing optimally in classification and has exhibited resilient and increased outcomes over the state-of-the-art techniques for the identification of respiratory illnesses in all the used datasets. The key contributions made by this study are summarized as follows:

\begin{figure*}[t]
    \centering
    \includegraphics[width=0.8\linewidth]{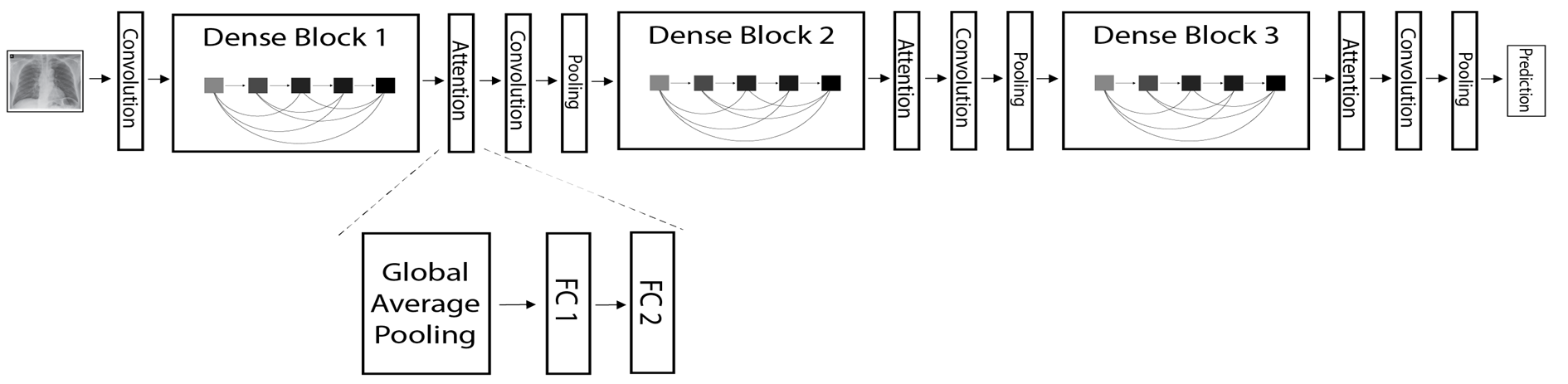}
    \caption{The Model Architecture}
    \label{fig:Model_Arch}
\end{figure*}

\begin{itemize}
    \item Propose and develop a novel model for X-ray images disease classification, enhancing the DenseNet121 model by introducing attention mechanism \cite{vaswani2017attention} and depthwise separable convolutions \cite{guo2019depthwise}, as shown in Figure~\ref{fig:Model_Arch}.
    \item Enhance the loss function of DenseNet121 based on focal loss \cite{pereira2020covid}.
    \item Conduct thorough experimental evaluations to demonstrate the effectiveness of the enhanced model using diverse X-ray image datasets with multiple chest diseases (viral Pneumonia, lung opacity, COVID-19). 
    \item Conduct a comparative analysis of the proposed model’s performance against other popular models such as DenseNet121, ResNet50, VGG16, and VGG19.
\end{itemize}


\section{Literature Review}

Extensive research has been conducted on the utilization of deep learning techniques for diagnosing respiratory diseases in patients using X-ray images. Below, we provide a concise overview of the most significant methodologies. 

In \cite{malik2023novel}, H. Malik et al. introduce a unique fusion model that combines hand-crafted features with Deep Convolution Neural Networks (DCNNs) to classify ten different chest diseases using chest X-ray images. This approach significantly enhances classification accuracy, demonstrating a novel contribution to automated diagnostic tools in medical imaging. The authors divided 181,719 X-ray images into segments of 80\%, 10\%, and 10\% for training, validation, and testing, respectively. The proposed fusion model in the paper achieved a classification accuracy of 98.20\% on the test set. 

The paper \cite{xu2022convolution} adopts a combination of transfer learning and network freezing that fastens the train processes and model performance that is used in the classification of respiratory disease using X-ray images. They use a 3-step process. First, they integrate a coordinate attention mechanism into the VGG16 network to create the VGG16-CoordAttention network. Second, the impact of freezing the network is examined using four sample models, including their model. Third and last, a 5-fold cross-validation is done to determine model efficiency. They got superior performance measures, including an accuracy of 92.73\% and an AUC of 97.71\%, higher than previous approaches. These results reflect the efficiency of coordinating attention combined with CNNs.

In \cite{kwon2023breathing}, a new region, referred to as the Breathing-Associated Facial Region, is proposed, which allows the breathing signal to be extracted at any angle using thermal imagery to investigate clinical diseases with nasal obstruction. The presented work significantly adds to the accuracy and reliability of non-contact measurements in the clinic. Setting. The authors divided the 5,232 samples in training, validation, and testing in a ratio of 89\%, 1\%, and 10\%, respectively. The proposed model demonstrated a notable average breathing accuracy of 90.9\%. 

In another study presented in \cite{rajpurohit2023improved}, the authors thoroughly compared the CNN models with cross-entropy loss, focal loss, and proposed hybrid loss. As shown in the results, the CNN model with hybrid loss outperformed other configurations with a higher accuracy and F1 score. This, in essence, underscores the effectiveness of the hybrid loss function in enhancing diagnostic capabilities for pneumonia detection.

By adopting the Faster R-CNN framework integrating the RoI Align versus the RoI, as in the paper \cite{ma2023effective}, pooling will be necessary. This change provides a better alignment of the extracted features with input images. Improving the detection of tuberculosis indicators on chest X-rays. It evaluates the performance standard of the model in the TBX11K dataset according to accuracy, precision, recall, and F1 score. The result shows improved accuracy in tuberculosis detection compared with the baseline Faster R-CNN model.

In the paper \cite{alshmrani2023deep}, the authors utilized the pre-trained VGG19 followed by CNN for feature extractions and a fully connected network for classifications. Their contribution is the multi-class classification of five lung diseases using X-ray images. The proposed architecture outperforms existing work with high accuracy (96.48\%), precision (97.56\%), recall (93.75\%), F1 score (95.62\%), and AUC (99.82\%). Like Inception V3, together augmentation and batch normalization are used in the paper \cite{prasher2023inception} for efficient binary classifications of X-ray images in tuberculosis and healthy cases. The accuracy rate shown by the authors, which is outstanding at 99\%, thus testifies to its viable application in tuberculosis detection.

Michail Mamalakis et al. developed a new deep-learning network named DenResCov-19 in their paper \cite{mamalakis2021denrescov} for robust and accurate binary and multi-class classification of various respiratory diseases (COVID-19, pneumonia, TB, and health). The authors use heterogeneous X-ray image datasets in their experiments. Moreover, the Monte Carlo cross-validation technique is used in their work to split the datasets for training, validation, and testing. Their work achieved promising results on different types of X-ray image datasets. DenResCov-19 network outperforms the prior works it seeks to improve, specifically the DenseNet-121 and ResNet-50 networks.
    
After deep analysis of all the prior studies on the classification of chest diseases using vision-based deep learning models, we identified the research gaps and the limitations as follows: i) imbalance in the available dataset, ii) lack of attention mechanisms,  iii) high computational cost is reducing applicability in real-time applications, iv) focus on binary classification, which makes the model very limited, and v) risk of overfitting because of small or specific datasets.

In this work, we focus on addressing the challenges that we mentioned. So, we chose DenseNet121 as a backbone and added depth-wise convolution to reduce the parameters to make the model lighter, which makes our model a suitable choice for real-world applications. Furthermore, we used focal loss to deal with the imbalance dataset issue. Also, we chose a dataset which has diversity in the classes to overcome the overfitting problem. Finally, we inserted an attention mechanism to highlight the most relevant information of input data while suppressing less important information.

\section{Proposed Method}

The human respiratory system could be subject to a range of pulmonary diseases. These include COVID, viral Pneumonia, and lung opacity, among other conditions. These disorders can have serious impacts on the human lungs; hence, deep learning-based methods speed up the diagnostic procedure and reduce time consumption for clinicians.

\begin{figure}[t]
    \centering
    \includegraphics[width=\columnwidth]{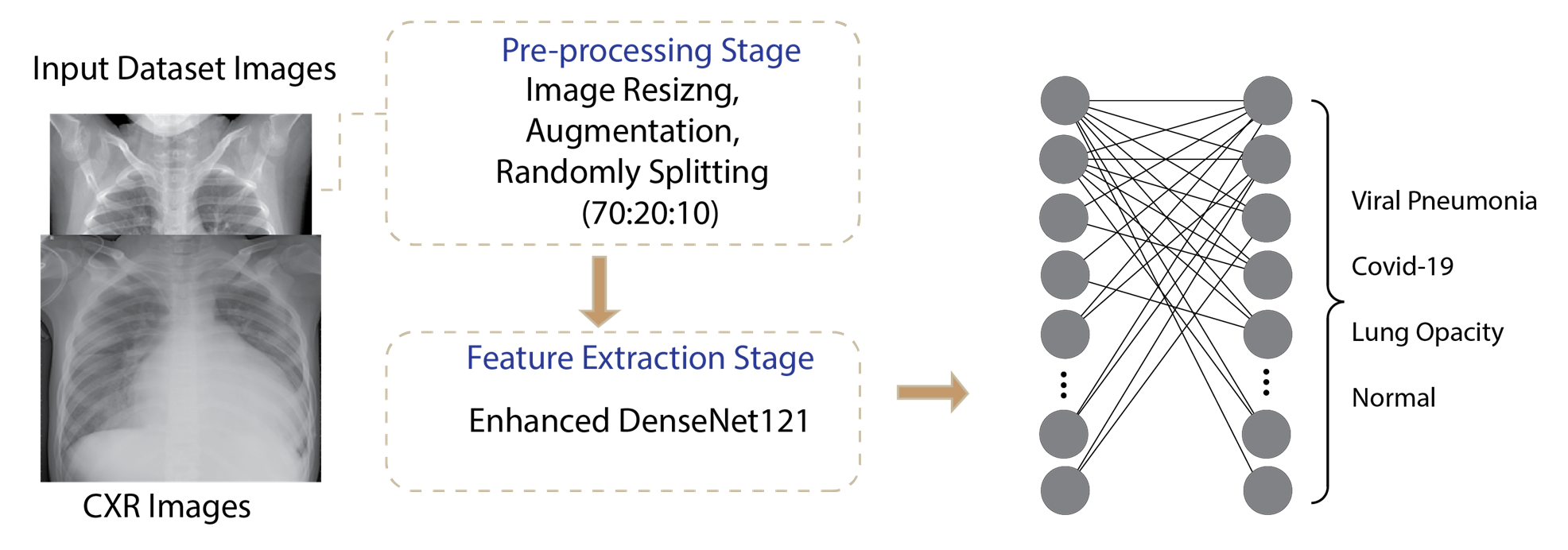}
    \caption{Proposed Methodology}
    \label{fig: Proposed Methodology}
\end{figure}

This work introduces a deep-learning classification model for the identification of prevalent respiratory disorders, including viral Pneumonia, Lung Opacity, and the more recent COVID-19. Figure~\ref{fig: Proposed Methodology} depicts the proposed method. The framework is comprised of four distinct phases: (i) utilization of X-ray images as inputs, (ii) pre-processing, (iii) feature extraction, and (iv) classification of the input image based on the presence of a disease.

In the initial stage, the X-ray images are used as inputs. Then, in the second stage, the images are subjected to pre-processing, which includes augmentation, resizing, and random splitting of data images into 70\% for training, 20\% for validation, and 10\% for testing. Subsequently, deep learning techniques are employed in both the third and fourth stages. The third phase is feature extraction, which the proposed model achieves. The picture categorization step utilizes the fully connected network. 

\subsection{Proposed Model}

DenseNet121 is a state-of-the-art CNN architecture that has demonstrated remarkable performance in image classification tasks \cite{vaswani2017attention}. DenseNet121 has two major aspects that make it stand out from other CNN models. First, it features a dense block structure, where each layer is coupled to every other layer in a feed-forward method. Second, it employs bottleneck layers that assist in minimizing the number of parameters without diminishing the number of characteristics learnt by the network.

In this work, two enhancements to the basic DenseNet121 model were introduced. A detailed description of each enhancement and the motivation behind each is given below. 

\subsubsection{Introducing Attention Blocks}
Inspired by \cite{vaswani2017attention}, an attention block is added after every dense block, as depicted in Figure~\ref{fig:Model_Arch}. The attention mechanism calculates the attention weights for each channel of the input feature maps. These weights are then utilized for the feature maps using element-wise multiplication, allowing the network to emphasize important features dynamically while minimizing the impact of less relevant ones. The attention block consists of global average pooling followed by two fully connected layers and, finally, the multiplication of the original feature maps by the attention score, as shown in Figure~\ref{fig:Model_Arch}. The details of the attention block are given below:
\begin{itemize}
    \item \textbf{Global average pooling:} This pooling operation reduced the spatial dimensions of the feature maps into a single vector, thereby collecting global information.
    \item \textbf{Fully connected Layers:} Global information is propagated across fully linked layers in order to learn the significance of various features.
    \item \textbf{Sigmoid activation:} The job of sigmoid activation is to restrict the attention score between 0 and 	1 to indicate the importance of each feature. 
    \item \textbf{Multiplication:} Ultimately, a multiplication operation between the initial feature map and the attention scores is done to highlight important features and diminish less pertinent ones.
\end{itemize}

\subsubsection{Depthwise Separable Convolutions}
Inspired by \cite{guo2019depthwise}, we employ depthwise convolutions instead of the standard convolutions. Depthwise separable convolutions are a factorized form of conventional convolutions that decrease the number of parameters and calculations, resulting in more efficient models.

Depthwise separable convolution is a convolutional operation that consists of two distinct steps:

\begin{itemize}
    \item \textbf{Depthwise convolution:} Convolution is applied on a single channel at a time, unlike standard CNNs, which operate on all the $M$ channels. So here, the filters will be of size $D_k \times D_k \times 1$. For $M$ channels in the input, $M$ filters have to be used. The result will be of size $D_p \times D_p \times M$.
    \item \textbf{Pointwise convolution:} A $1 \times 1$ convolution operation is executed on the $M$ channels in a point-wise fashion. So, the filter size will be $1 \times 1 \times M$ for this operation. Say we apply $N$ such filters; the output size becomes $D_p \times D_p \times N$.
\end{itemize}

\textbf{Table~\ref{tab: Complexity comparison}} displays the contrast between the complexity of conventional and depthwise separable convolution.
\begin{table}[h]
    \centering
    \begin{tabular}{cc}
        \toprule
        \textbf{Type of Convolution} & \textbf{Complexity} \\ 
        \midrule
        Standard & $N \times D_p^2 \times D_k^2 \times M$ \\ 
        Depthwise separable & $M \times D_p^2 \times (D_k^2 + N)$ \\ 
        \bottomrule
    \end{tabular}
    \caption{Complexity comparison}
    \label{tab:complexity_comparison}
\end{table}
Where, \textbf{N} = Number of filters; \textbf{M} = Channels' number
\textbf{D\textsubscript{p}} = Width/Height of output feature map;
and \textbf{D\textsubscript{k}} = Width/Height of filter

\section{Experimental Setup}

The experiment setup for the proposed model includes multiple details regarding the hardware configuration, dataset sources, and hyper-parameters used in the enhanced DenseNet121 model. The hardware utilized compromises NVIDIA GeForce GTX 106 with \textbf{6} GB memory, and the RAM size is \textbf{16} GB. The dataset\cite{rahman2021exploring}\cite{chowdhury2020can} used contains \textbf{21,265} X-ray images with four classes as shown in {Table~\ref{tab:description_of_the_dataset}}. The hyper-parameters used in our experiments are shown in {Table~\ref{tab:hyper-parameters}}. 

\begin{table}[h]
    \centering
    \begin{tabular}{ccccc}
        \toprule
        \textbf{Exp.} & \textbf{Batch Size} & \textbf{LR} & \textbf{Epochs} & \textbf{Optimizer} \\ 
        \midrule
        1 & 64 & 0.001 & 20 & Adam \\ 
        2 & 128 & 0.001 & 100 & Adam \\ 
        \bottomrule
    \end{tabular}
    \caption{Hyper-parameters}
    \label{tab:hyper-parameters}
\end{table}
\subsection{Dataset}
The dataset selected in this study consists of samples from cases of COVID-19, lung opacity, viral Pneumonia, and healthy lungs. It was acquired from Kaggle, an online resource of open-source datasets\cite{rahman2021exploring}\cite{chowdhury2020can}. The datasets consist of a total of 21,265 labelled X-ray images, of which 3,716 were COVID-19, 6,012 were lung opacity, 10,192 were viral Pneumonia, and 1,345 were normal lungs. We split the collected data into 70\% for training, 10\% for validation, and 20\% for testing. Table~\ref{tab:description_of_the_dataset} illustrates the dataset. 
\begin{table}[h]
    \centering
    \begin{tabular}{p{1cm}p{1cm}ccc}
        \toprule
        \textbf{Disease} & \textbf{Train} & \textbf{Validation} & \textbf{Testing} & \textbf{Total} \\ 
        \midrule
        COVID-19 & 2,531 & 461 & 724 & 3,716 \\ 
        Lung Opacity & 4,208 & 601 & 1,203 & 6,012 \\ 
        Normal & 7,134 & 1,019 & 2,039 & 10,192 \\ 
        Viral Pneumonia & 941 & 134 & 270 & 1,345 \\ 
        Total & 14,814 & 2,115 & 4,236 & 21,265 \\ 
        \bottomrule
    \end{tabular}
    \caption{Description of the dataset}
    \label{tab:description_of_the_dataset}
\end{table}

\begin{figure}[]
\centering
\includegraphics[width=\linewidth]{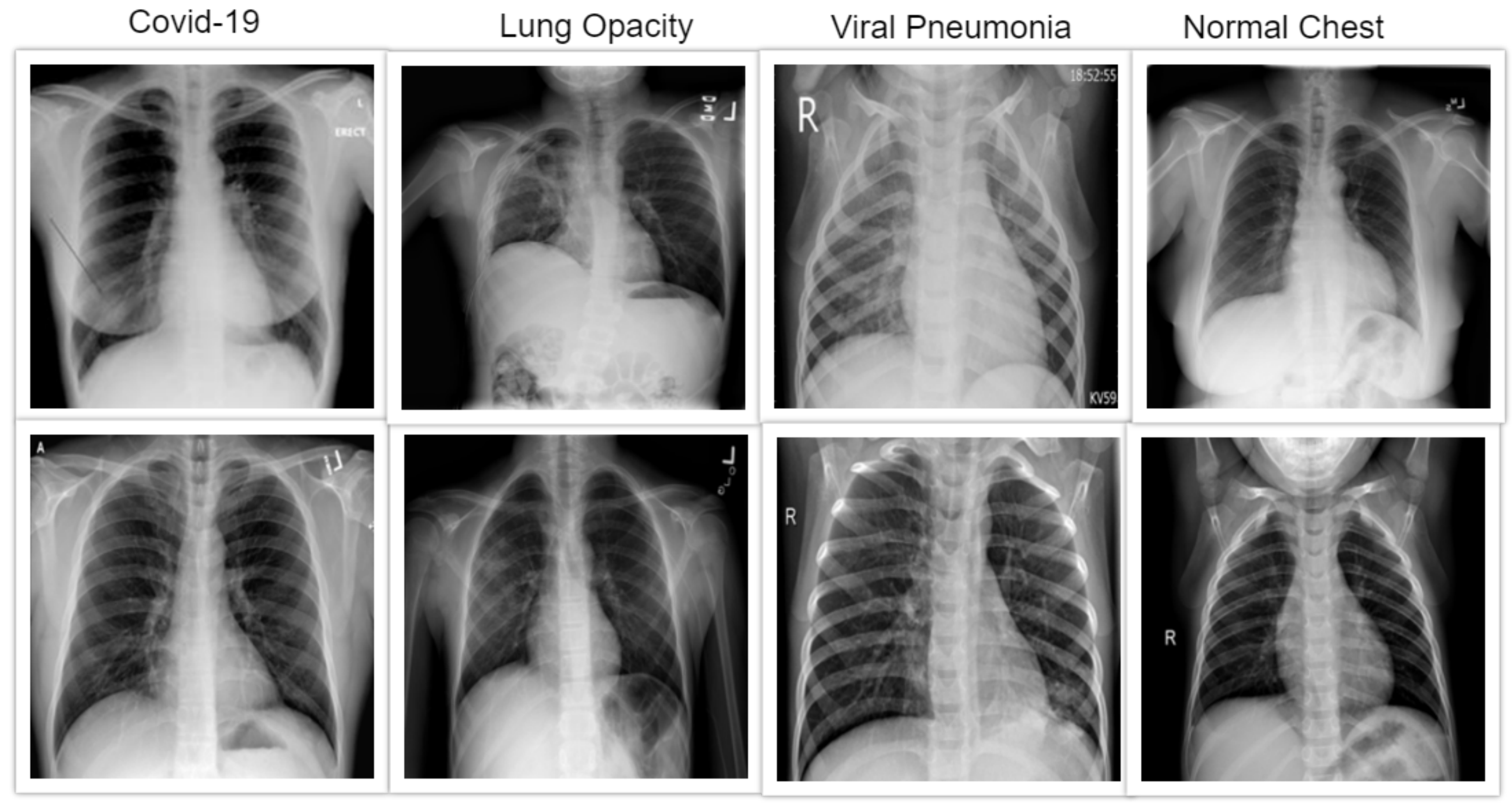}
\caption{A sample of chest diseases from X-ray images dataset. The size of all images is \textbf{229x299}.}
\label{fig:sample_images}
\end{figure}

\section{Results and Discussions }

While utilizing all the elements in our experimental setup, we were able to implement an enhanced model version of DenseNet121. The proposed model outperforms the original DenseNet121, ResNet50, VGG16, and VGG19 on the same dataset. The precision, recall, accuracy, and loss after each epoch were recorded. Also, Focal loss as a loss function instead of cross-entropy has been utilized in our model, which helps the model solve the problem of class imbalance. Also, our proposed model outperformed the DenseNet121 in \cite{constantinou2023covid}, where they used the same dataset for the same purpose.

\subsection{Performance of Enhanced DenseNet}
\begin{table*}[]
    \centering
    \renewcommand{\arraystretch}{1.5}  
    \resizebox{0.9\textwidth}{!}{%
        \begin{tabular}{cccccc}
            \toprule
            \textbf{Model} & \textbf{Accuracy} & \textbf{Precision} & \textbf{Recall} & \textbf{No. of Parameters} & \textbf{Inference Time} \\ 
            \midrule
            Enhanced DenseNet-Focal & 91.96 & 93.01 & 91.06 & 6,437,380 & 0.201 sec \\ 
            Enhanced-DenseNet & 91.90 & 92.32 & 91.55 & 6,437,380 & 0.304 sec \\ 
            DenseNet121 & 87.86 & 88.00 & 87.20 & 6,957,956 & 0.232 sec \\ 
            ResNet50 & 88.32 & 88.30 & 88.31 & 25,557,032 & 0.272 sec \\ 
            VGG16 & 79.84 & 79.67 & 79.84 & 138,357,544 & 0.703 sec \\ 
            VGG19 & 78.49 & 78.32 & 78.49 & 143,667,240 & 0.762 sec \\ 
            \bottomrule
        \end{tabular}%
        } 
    \caption{Summary of Comparisons (20 epochs)}
    \label{tab:SummaryOfComparisons}
\end{table*}

The accuracies of our models with and without focal loss with (20 epochs) in experiment 1 are \textbf{91.96\%} and \textbf{91.90\%}, respectively, while the precisions were \textbf{93.01\%} and \textbf{92.32\%}, respectively. The recalls were \textbf{91.06\%} and \textbf{91.55\%}, respectively. Figure~\ref{fig: Performance of Enhanced DenseNet121 with Focal Loss (100 epochs)} indicates how the performance of the model is affected by the increase in the number of epochs. The performance of the model improves because the focal loss forces the model to pay attention to the minority parts of the X-ray images where the disease symptoms lie. For experiment 2 (100 epochs), the accuracy, precision and recall of our proposed model with a focal loss is \textbf{94.94\%}, \textbf{95.14\%} and \textbf{94.53\%}, respectively, as demonstrated in Figure \ref{fig: Performance of Enhanced DenseNet121 with Focal Loss (100 epochs)}.

\begin{table}[H]
    \centering
    \renewcommand{\arraystretch}{1}  
    \resizebox{\columnwidth}{!}{%
        \begin{tabular}{p{3cm}>{\centering\arraybackslash}p{1cm}>{\centering\arraybackslash}p{1cm}>{\centering\arraybackslash}p{1cm}}
            \toprule
            \textbf{Model} & \textbf{Accuracy} & \textbf{Precision} & \textbf{Recall} \\ 
            \midrule
            Enhanced DenseNet & 94.94 & 95.14 & 94.53 \\ 
            Enhanced-DenseNet with Focal-Loss \cite{constantinou2023covid} & 93 & 93 & 93 \\ 
            \bottomrule
        \end{tabular}
    }  
    \caption{Summary of Comparisons (100 epochs)}
    \label{tab:summary_of_comparisons_100_epochs}
    \vspace{-0.5cm}  
\end{table}

\subsection{Comparative Analysis}
We performed experiments to compare our model with the state-of-the-art models. Our model has the fewest number of parameters. Additionally, has the shortest inference time on Google Colab. In the first experiment, results were compared after 20 epochs of training. The result of this is shown in \textbf{Table \ref{tab:SummaryOfComparisons}}. In the second experiment, we evaluate the impact of using focal loss on our model. The comparison is shown in \textbf{Table ~\ref{tab: Summary of Comparisons (100 epochs)}}.  In Figures 8-12, we plot several performance metrics of the considered models against the number of epochs. This also serves to shed light on the convergence characteristics of the training process.

In order to compare the interpretability capacity of our model and the state-of-the-art models, we employ Grad-Cam \cite{8237336} for heatmap visualizations on selected X-ray images. 
We show on a side-by-side basis how our model performs with superiority in Figures 4-7. We also observe that the addition
of the attention block has a crucial role in focusing on the most relevant features in the images, which significantly assists the model during classification.

\begin{figure}[H]  
    \centering
    \begin{minipage}{0.3\columnwidth}  
        \centering
        \includegraphics[width=\linewidth]{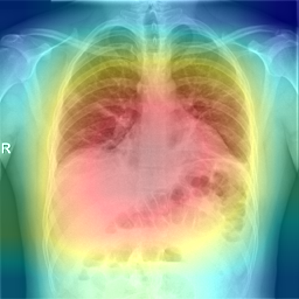}
        \label{fig:DenseNet_correct}
    \end{minipage}
    \hspace{0.08\linewidth}  
    \begin{minipage}{0.3\columnwidth}  
        \centering
        \includegraphics[width=\linewidth]{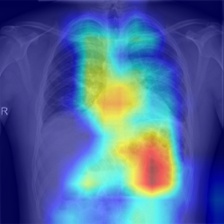}
        \label{fig:DenseNet_wrong}
    \end{minipage}    
    \caption{Enhanced DenseNet121 on the left-hand side, and original DenseNet121 on the right-hand side.}
    \label{fig:DenseNet_Classification_Comparison}
\vspace{-0.5cm}  
\end{figure}

\begin{figure}[H]  
    \centering
    \begin{minipage}{0.3\columnwidth}  
        \centering
        \includegraphics[width=\linewidth]{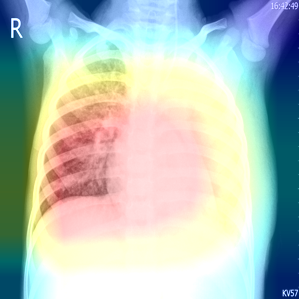}
        \label{fig:ResNet_correct}
    \end{minipage}
    \hspace{0.08\linewidth}  
    \begin{minipage}{0.3\columnwidth}  
        \centering
        \includegraphics[width=\linewidth]{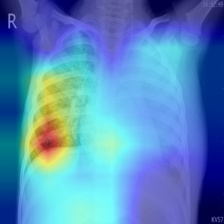}
        \label{fig:ResNet_wrong}
    \end{minipage}    
    \caption{Enhanced DenseNet121 on the left-hand side, and ResNet50 on the right-hand side.}
    \label{fig:ResNet_Classification_Comparison}
\vspace{-0.5cm}  
\end{figure}

\begin{figure}[H]  
    \centering
    \begin{minipage}{0.3\columnwidth}  
        \centering
        \includegraphics[width=\linewidth]{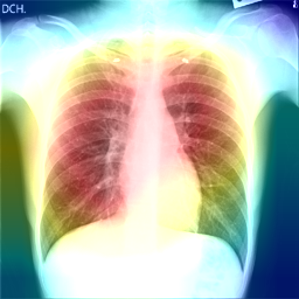}
        \label{fig:vgg16_correct}
    \end{minipage}
    \hspace{0.08\linewidth}  
    \begin{minipage}{0.3\columnwidth}  
        \centering
        \includegraphics[width=\linewidth]{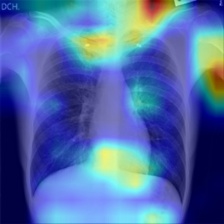}
        \label{fig:vgg16_wrong}
    \end{minipage}    
    \caption{Enhanced DenseNet121 on the left-hand side, and VGG16 on the right-hand side.}
    \label{fig:VGG16_Classification_Comparison}
\end{figure}

\vspace{-0.5cm}  

\begin{figure}[H]  
    \centering
    \begin{minipage}{0.3\columnwidth}  
        \centering
        \includegraphics[width=\linewidth]{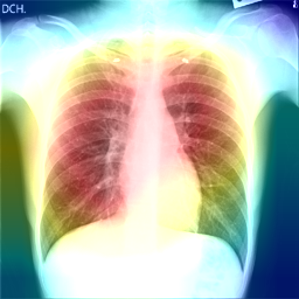}
        \label{fig:vgg19_correct}
    \end{minipage}
    \hspace{0.08\linewidth}  
    \begin{minipage}{0.3\columnwidth}
        \centering
        \includegraphics[width=\linewidth]{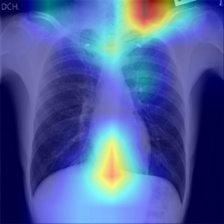}
        \label{fig:vgg19_wrong}
    \end{minipage}    
    \caption{Enhanced DenseNet121 on the left-hand side, and VGG19 on the right-hand side.}
    \label{fig:VGG19_Classification_Comparison}
\vspace{-0.5cm}
\end{figure}

\begin{figure*}[t!]
    \centering
    \begin{tabular}{cccc}
        \includegraphics[width=0.2\linewidth]{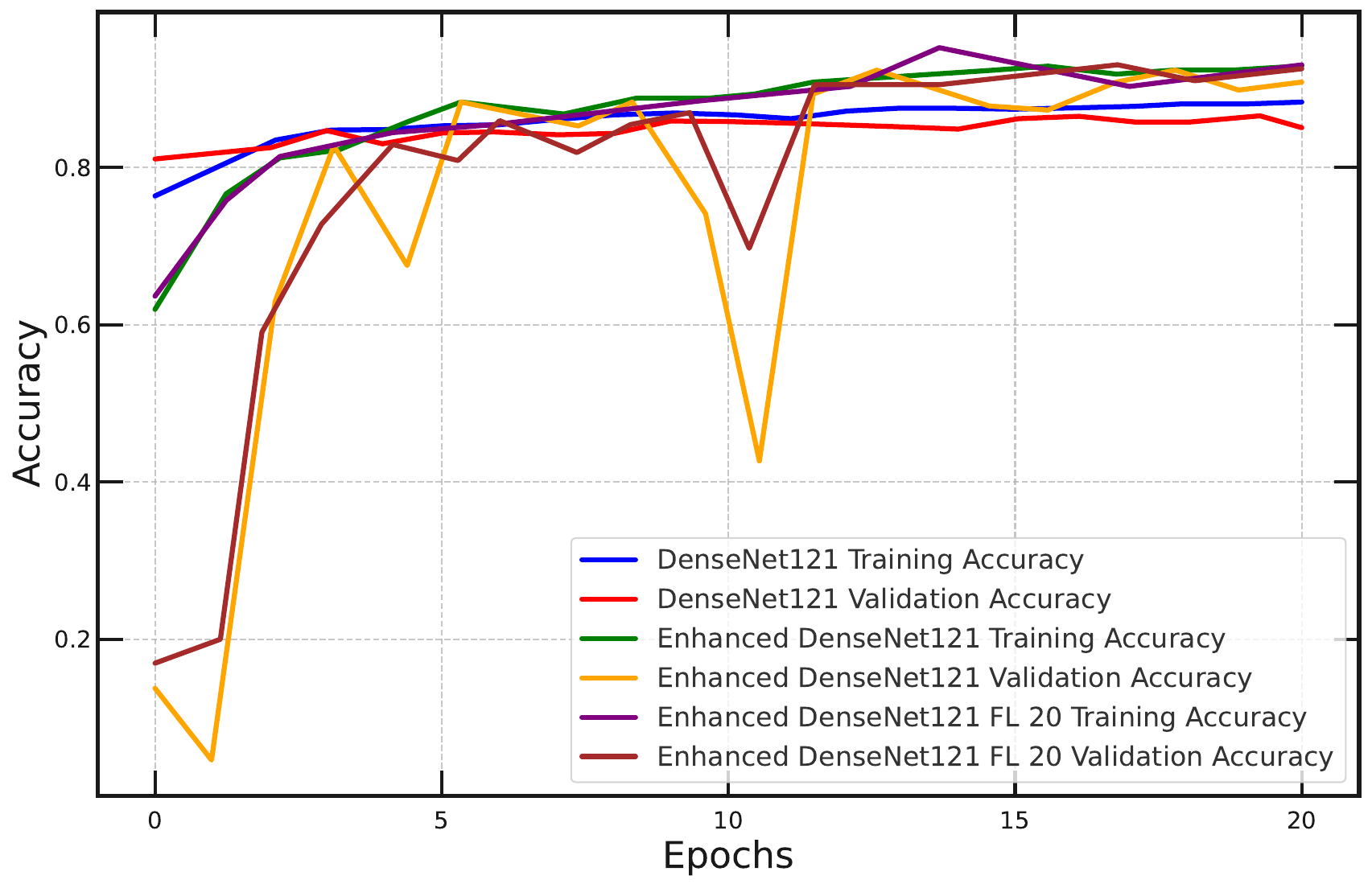} & 
        \includegraphics[width=0.2\linewidth]{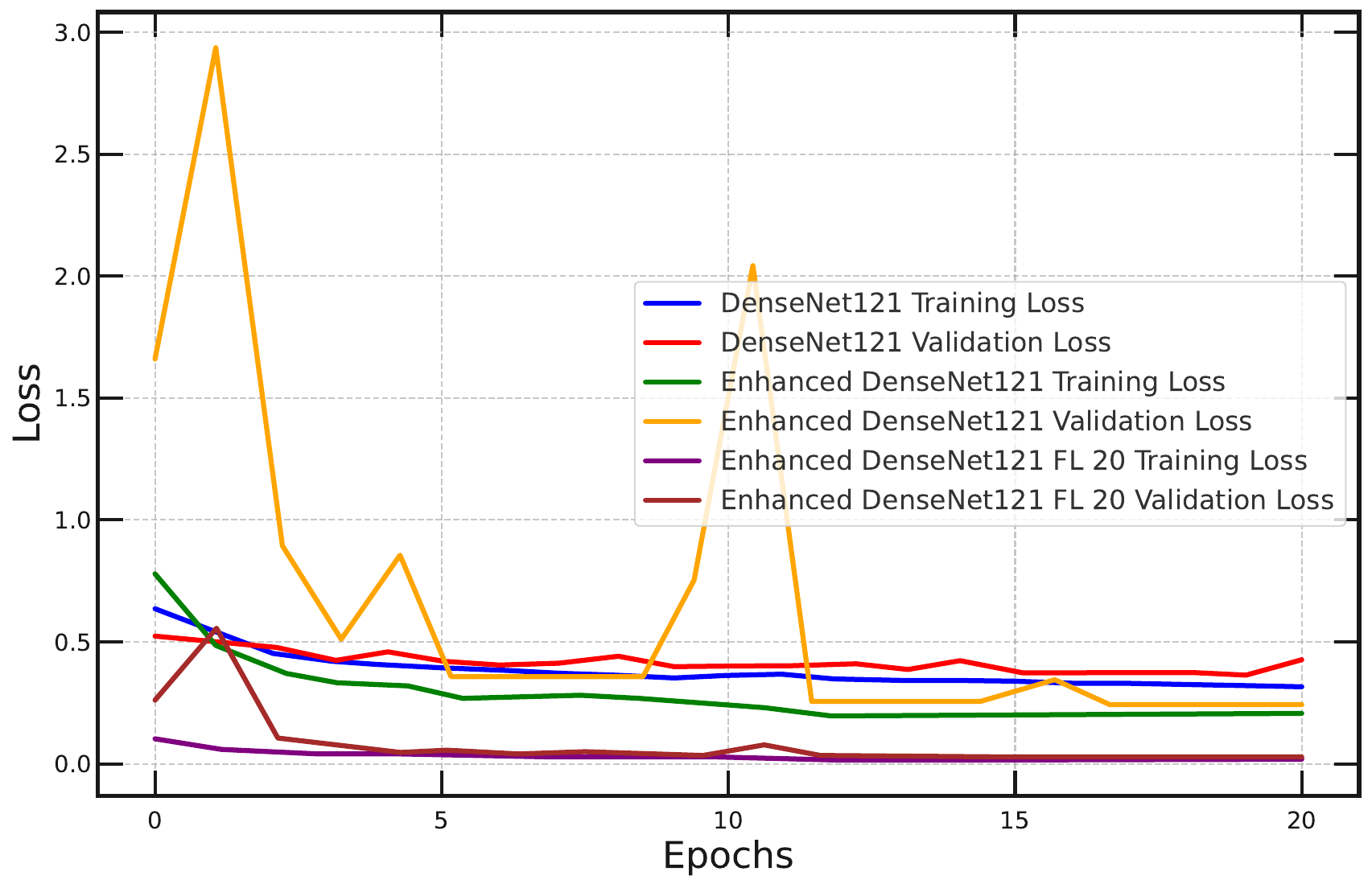} & 
        \includegraphics[width=0.2\linewidth]{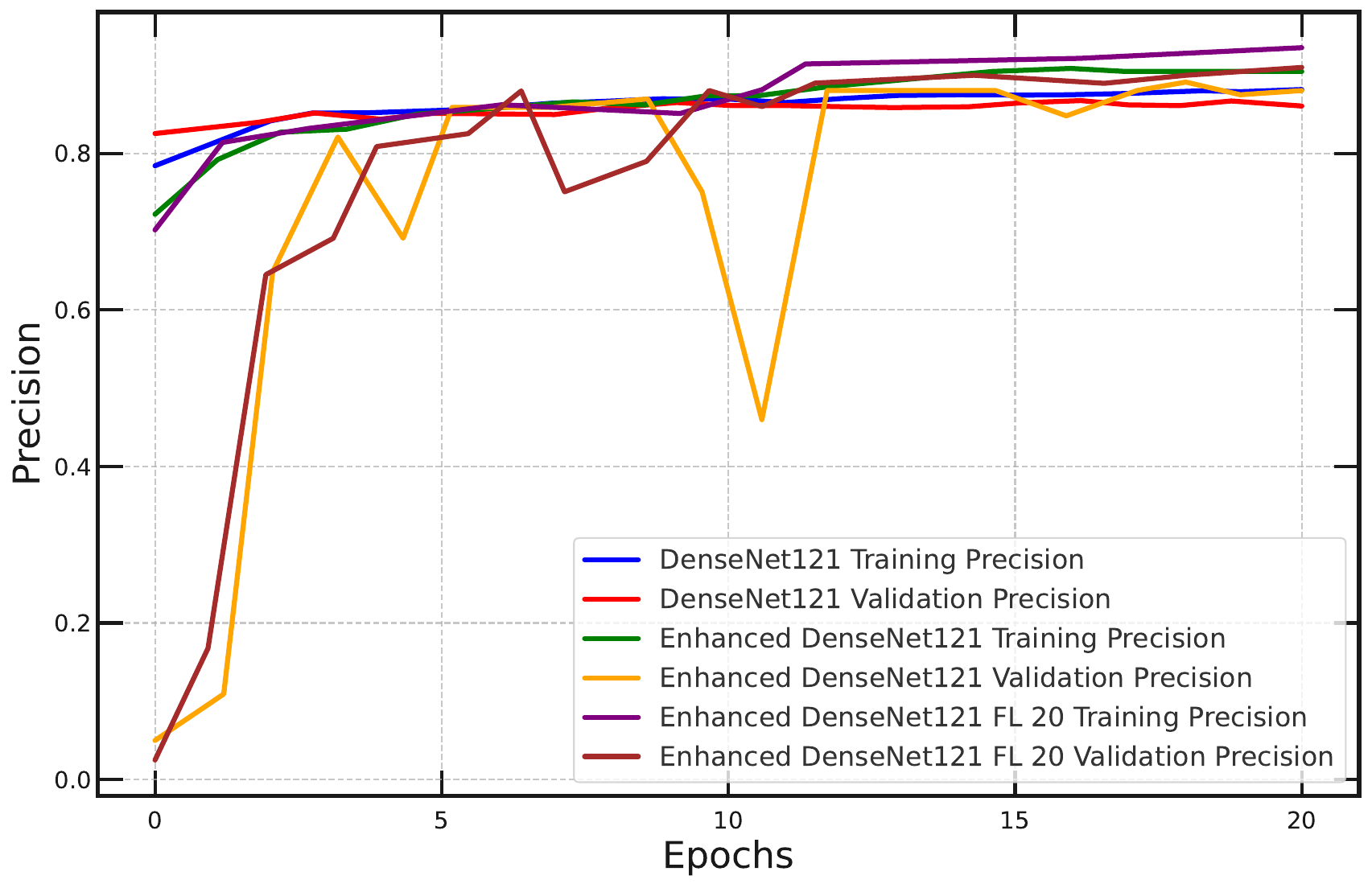} & 
        \includegraphics[width=0.2\linewidth]{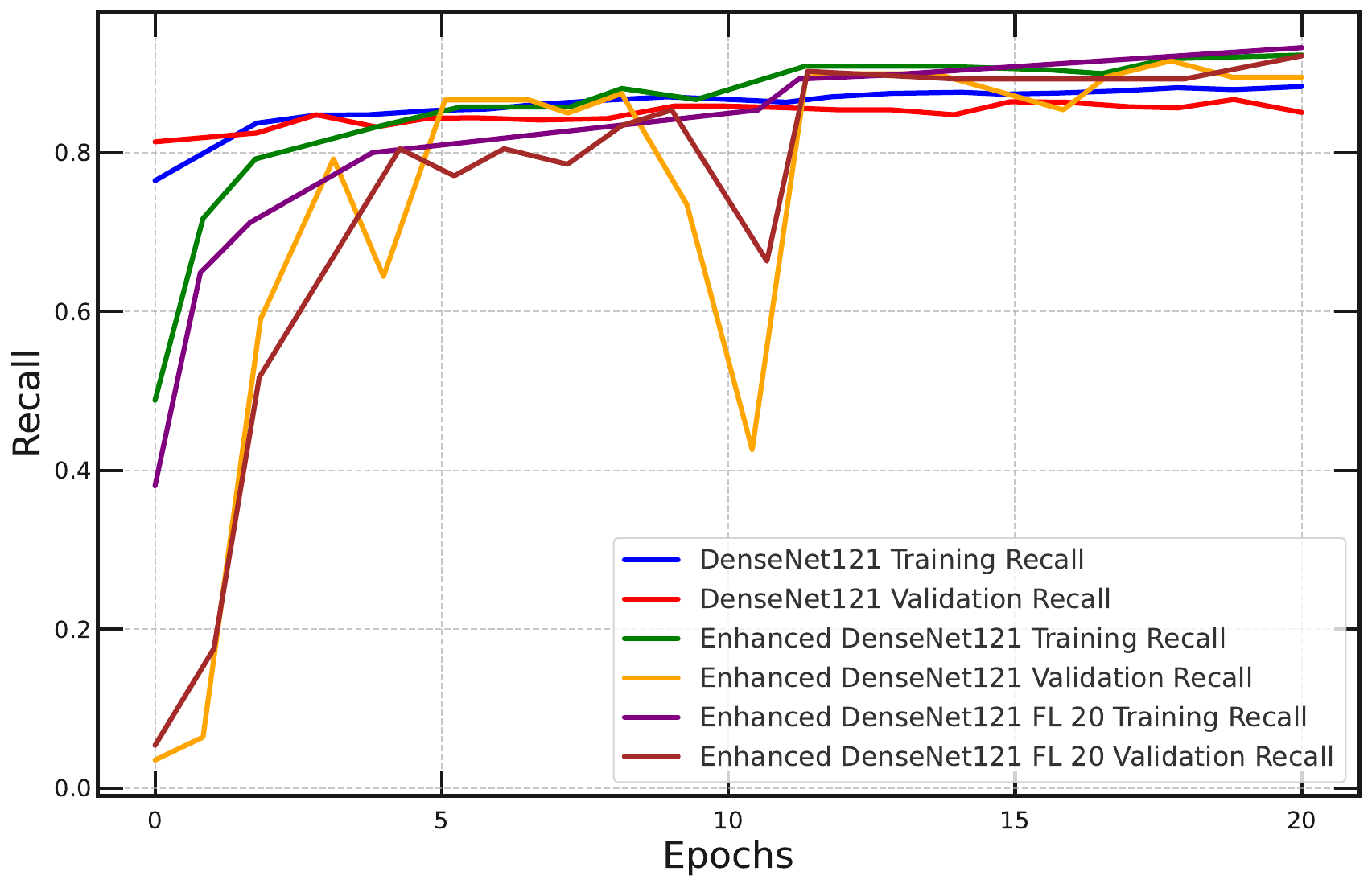} \\
        {(a) Accuracy Comparison} & {(b) Loss Comparison} & {(c) Precision Comparison} & {(d) Recall Comparison} 
    \end{tabular}
    \caption{Performance of original DenseNet121, enhanced DenseNet121 with/without Focal Loss}
    \label{fig:comp_summary}
\end{figure*}

\begin{figure}[H]
    \centering
    \includegraphics[width=0.2\textwidth]{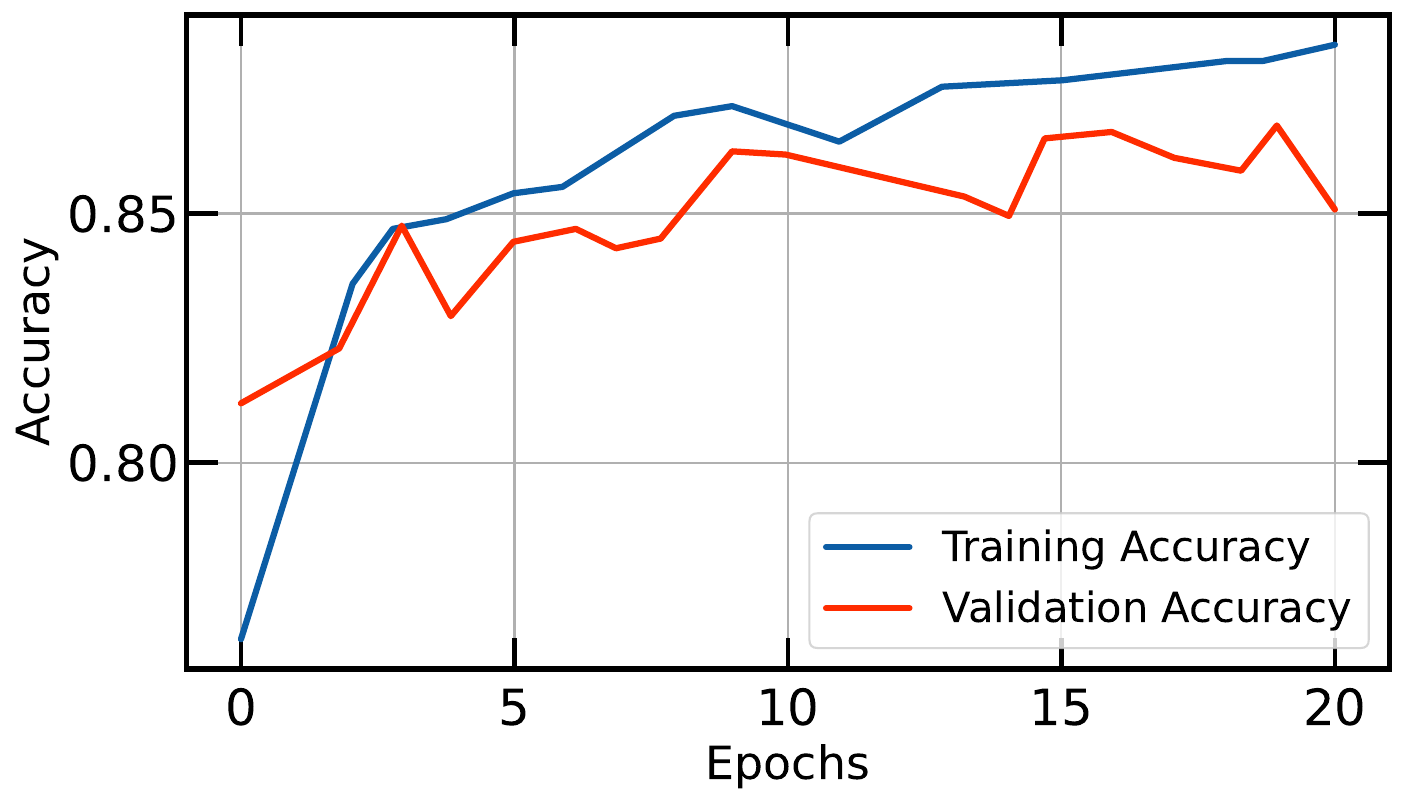}  
    \hfill
    \includegraphics[width=0.2\textwidth]{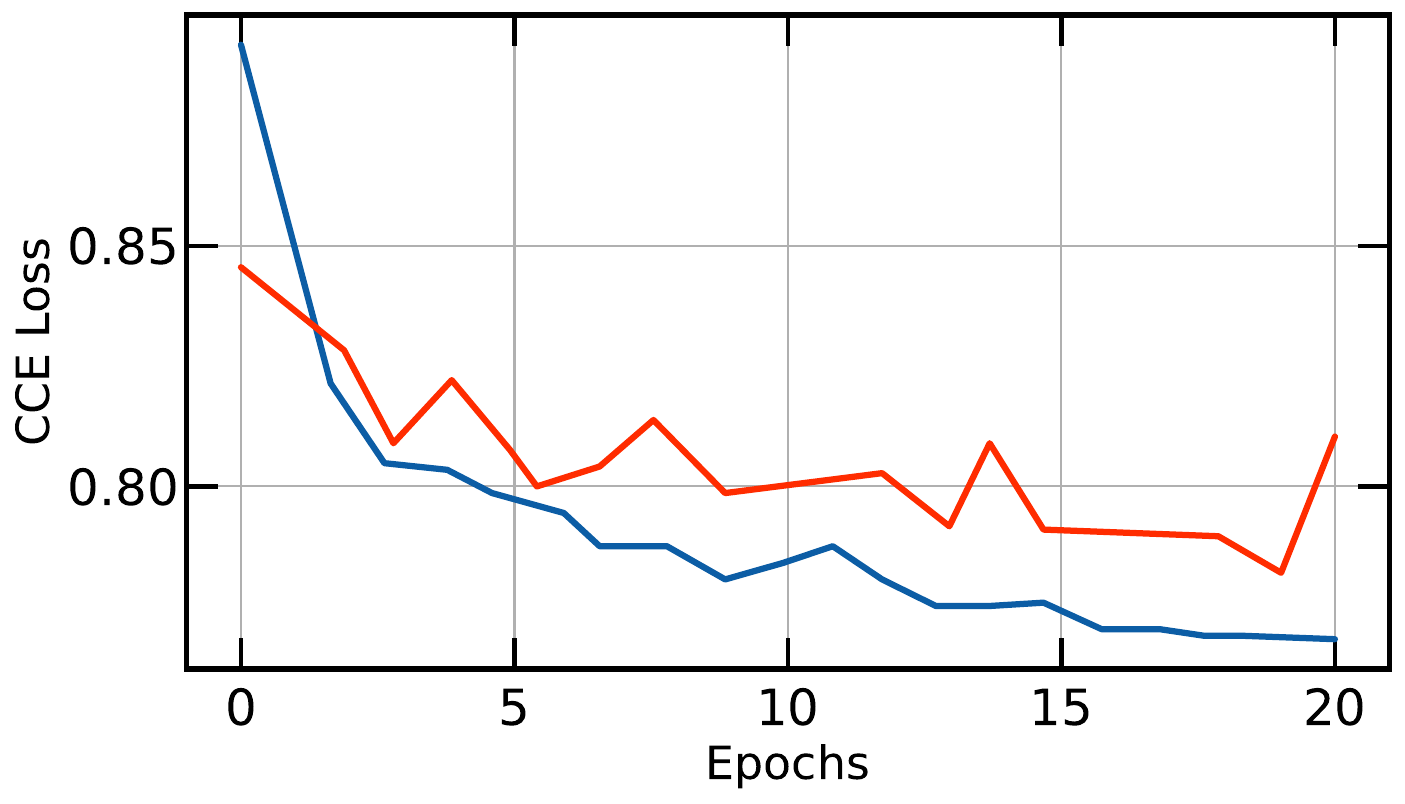}  
    \hfill
    \includegraphics[width=0.2\textwidth]{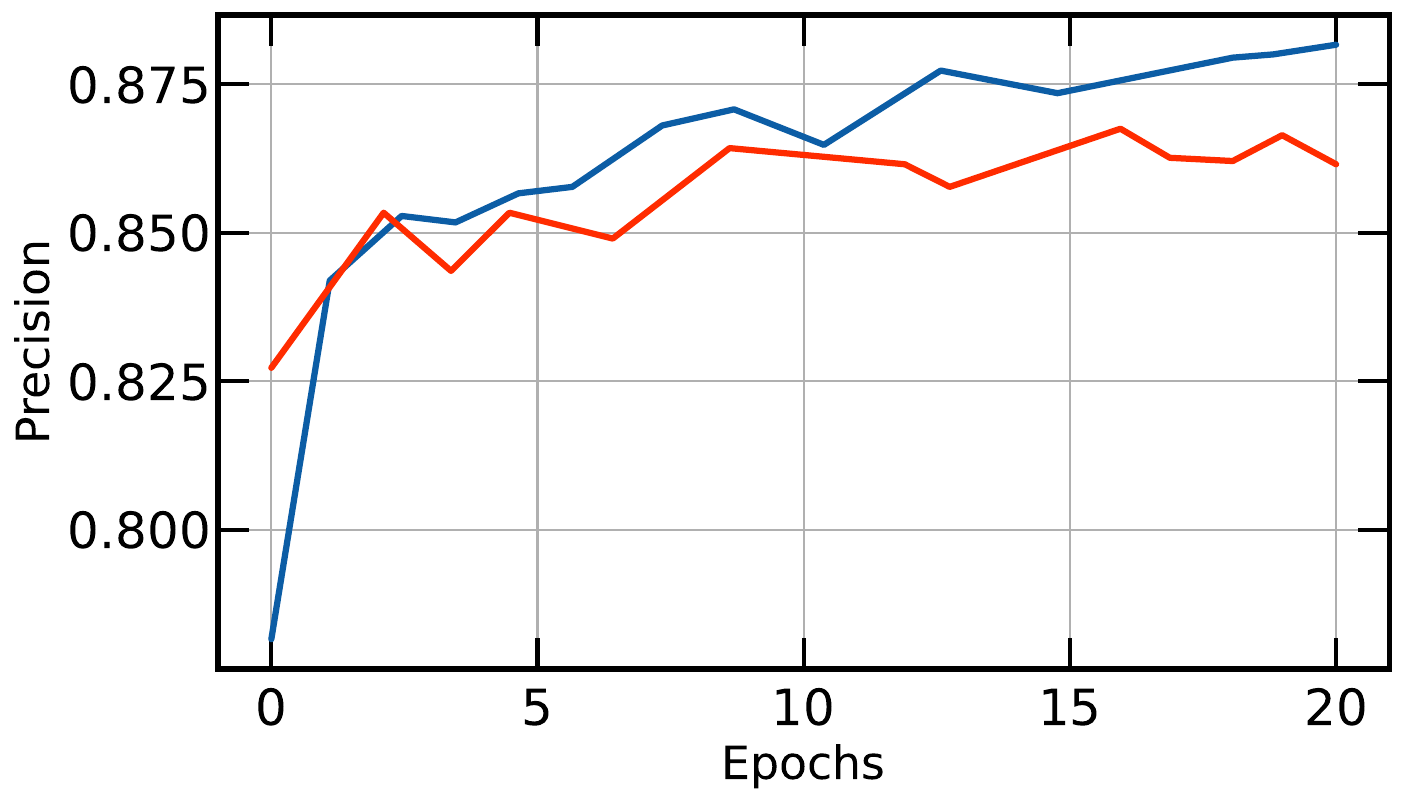}  
    \hfill
    \includegraphics[width=0.2\textwidth]{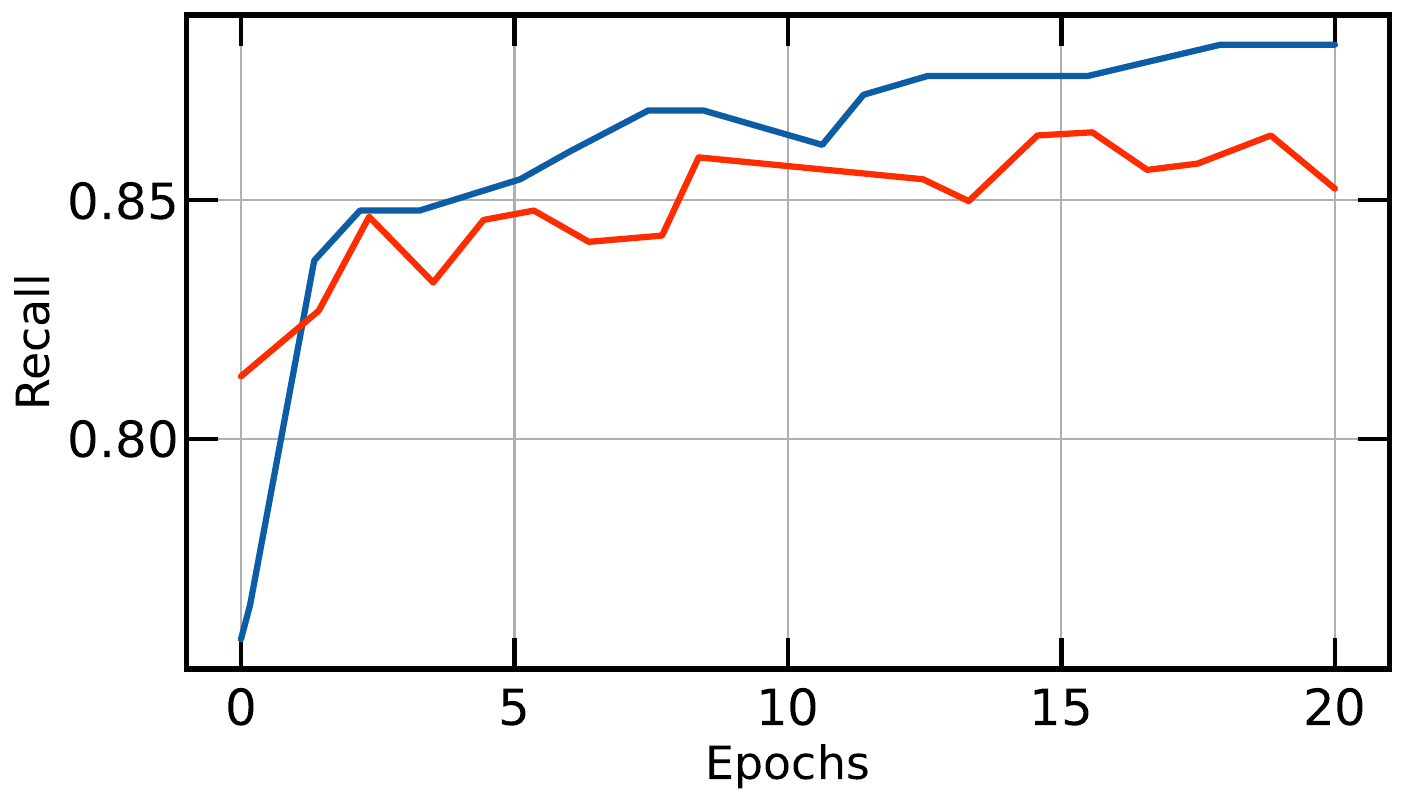}  
    \caption{Performance of ResNet50}
    \label{fig: Performance of ResNet50}
\vspace{-0.3cm}
\end{figure}

\begin{figure}[H]
    \centering
    \includegraphics[width=0.2\textwidth]{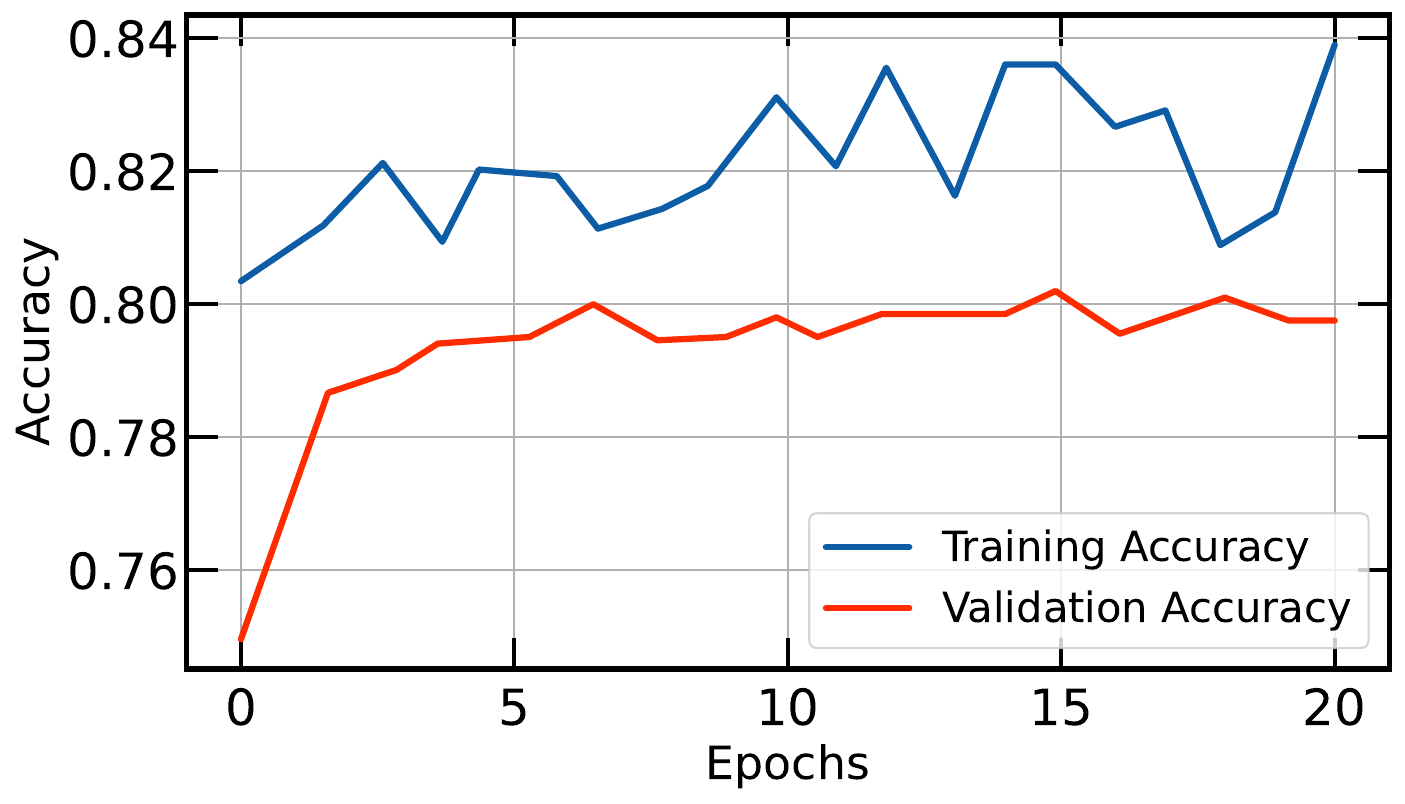}  
    \hfill
    \includegraphics[width=0.2\textwidth]{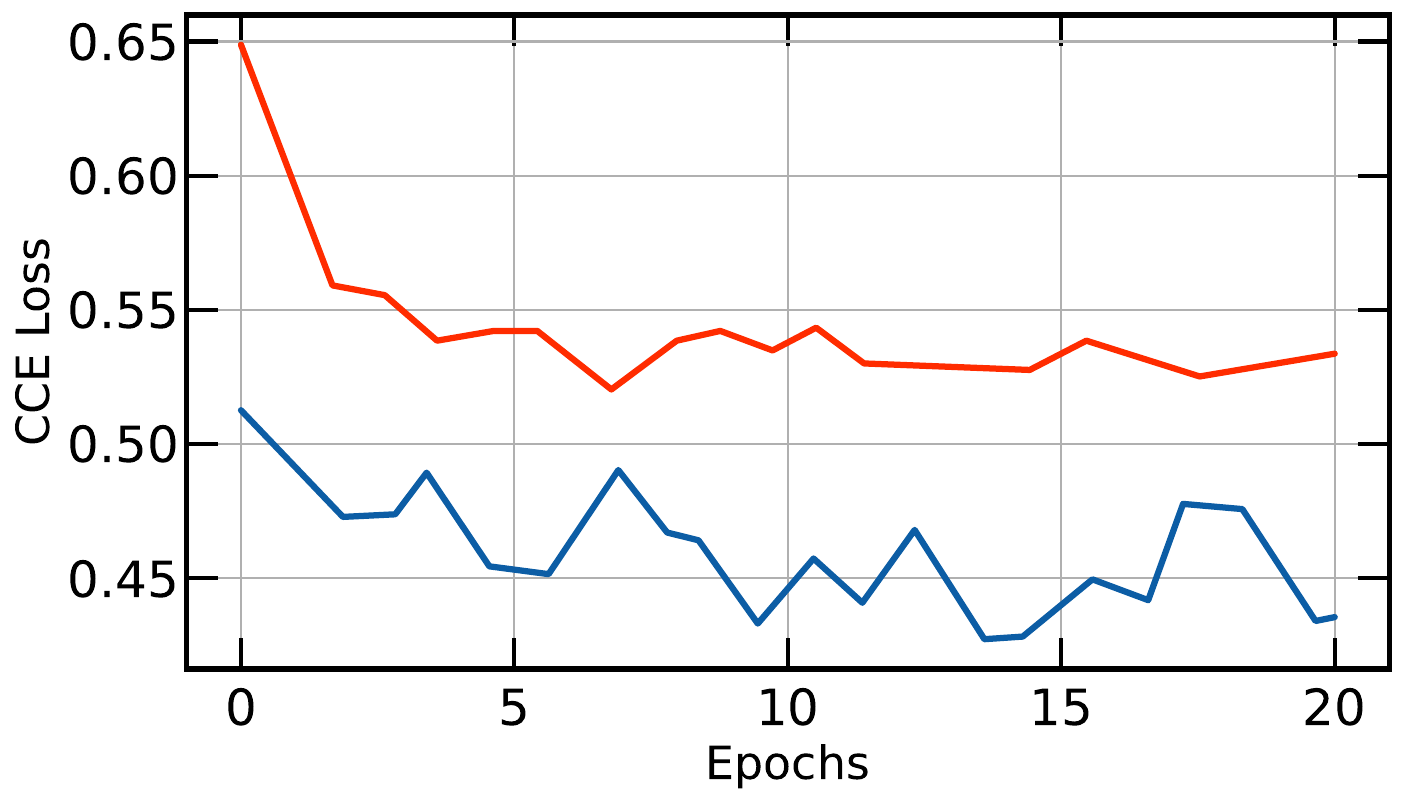}  
    \hfill
    \includegraphics[width=0.2\textwidth]{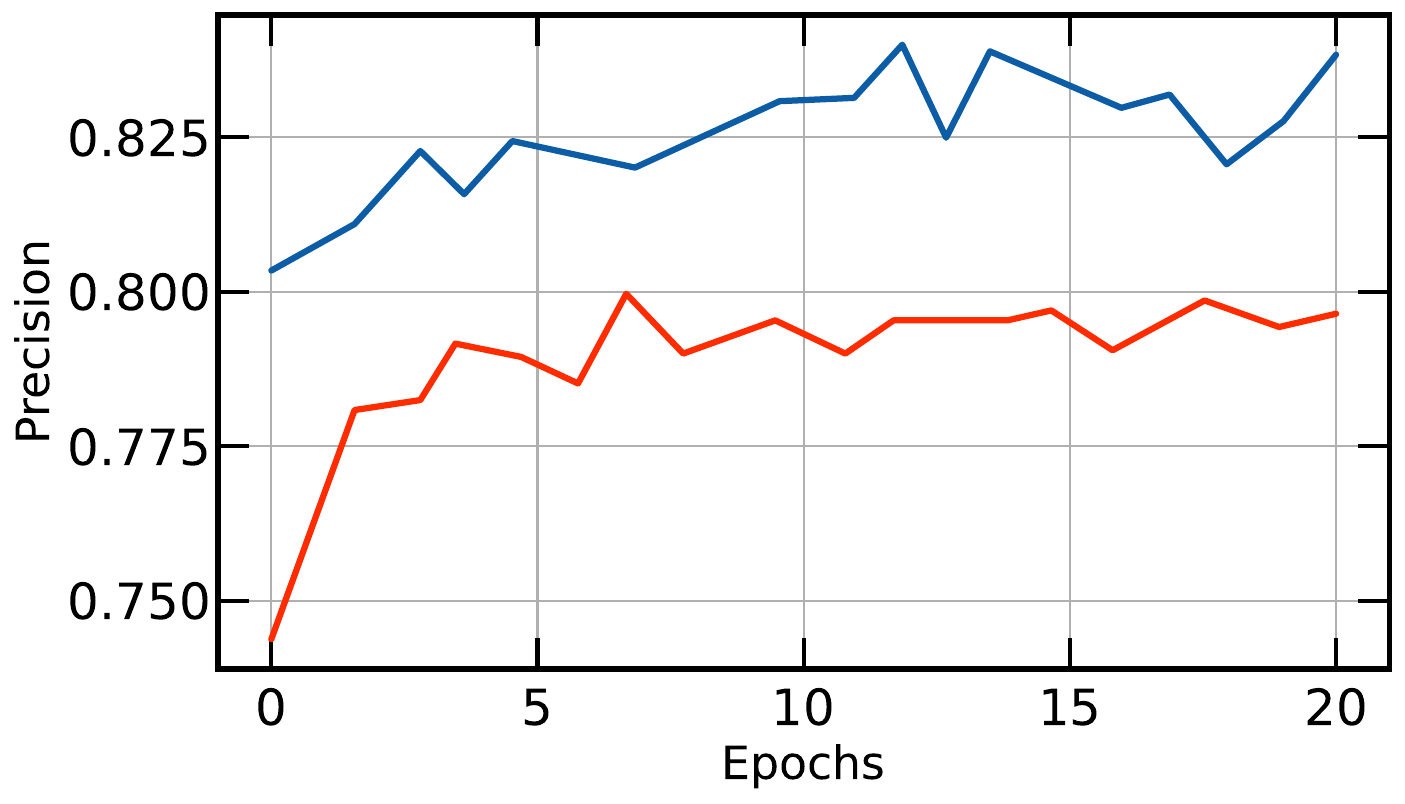}  
    \hfill
    \includegraphics[width=0.2\textwidth]{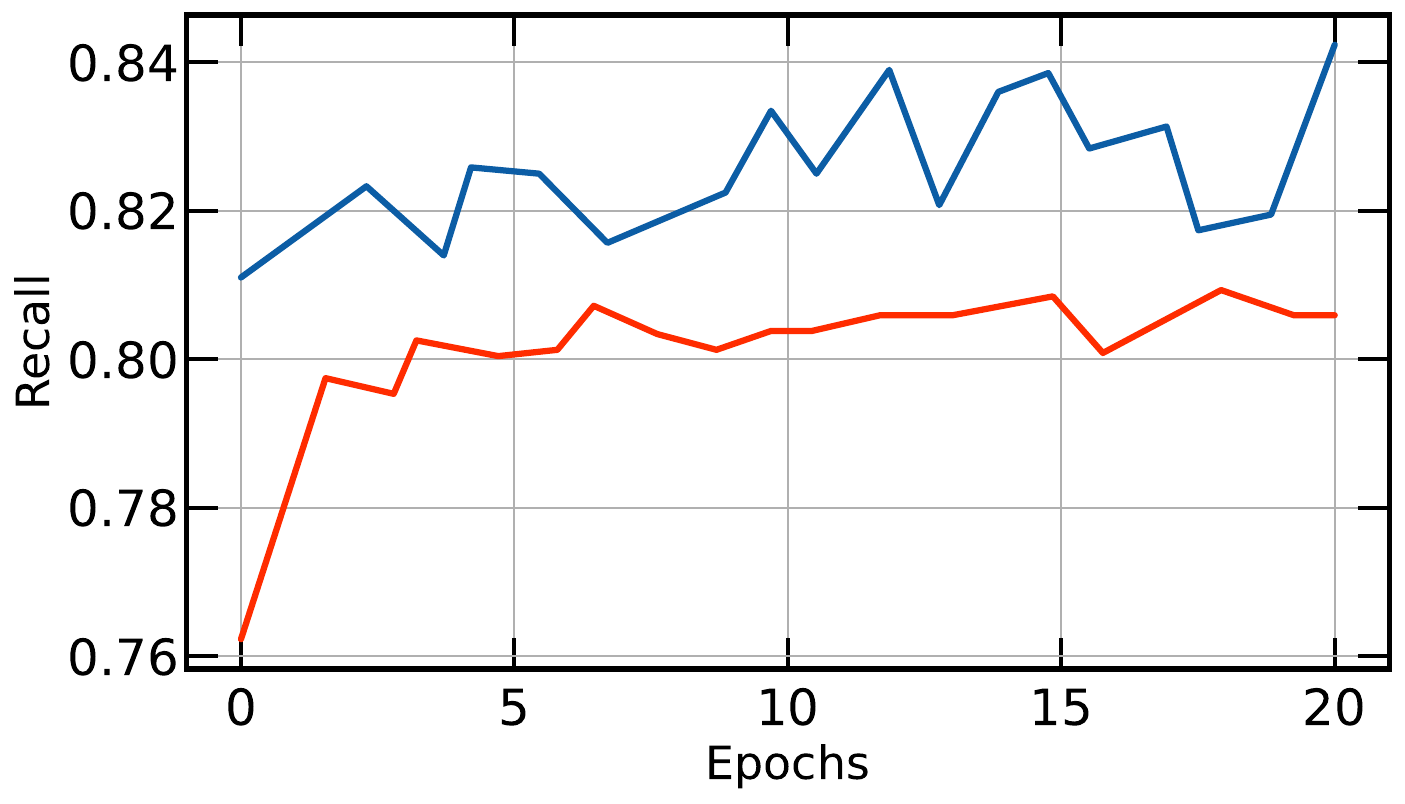}  
    \caption{Performance of VGG16}
    \label{fig: Performance_of_VGG16}
\vspace{-0.3cm}
\end{figure}

\vspace{-0.3cm}

\begin{figure}[H]
    \centering
    \includegraphics[width=0.2\textwidth]{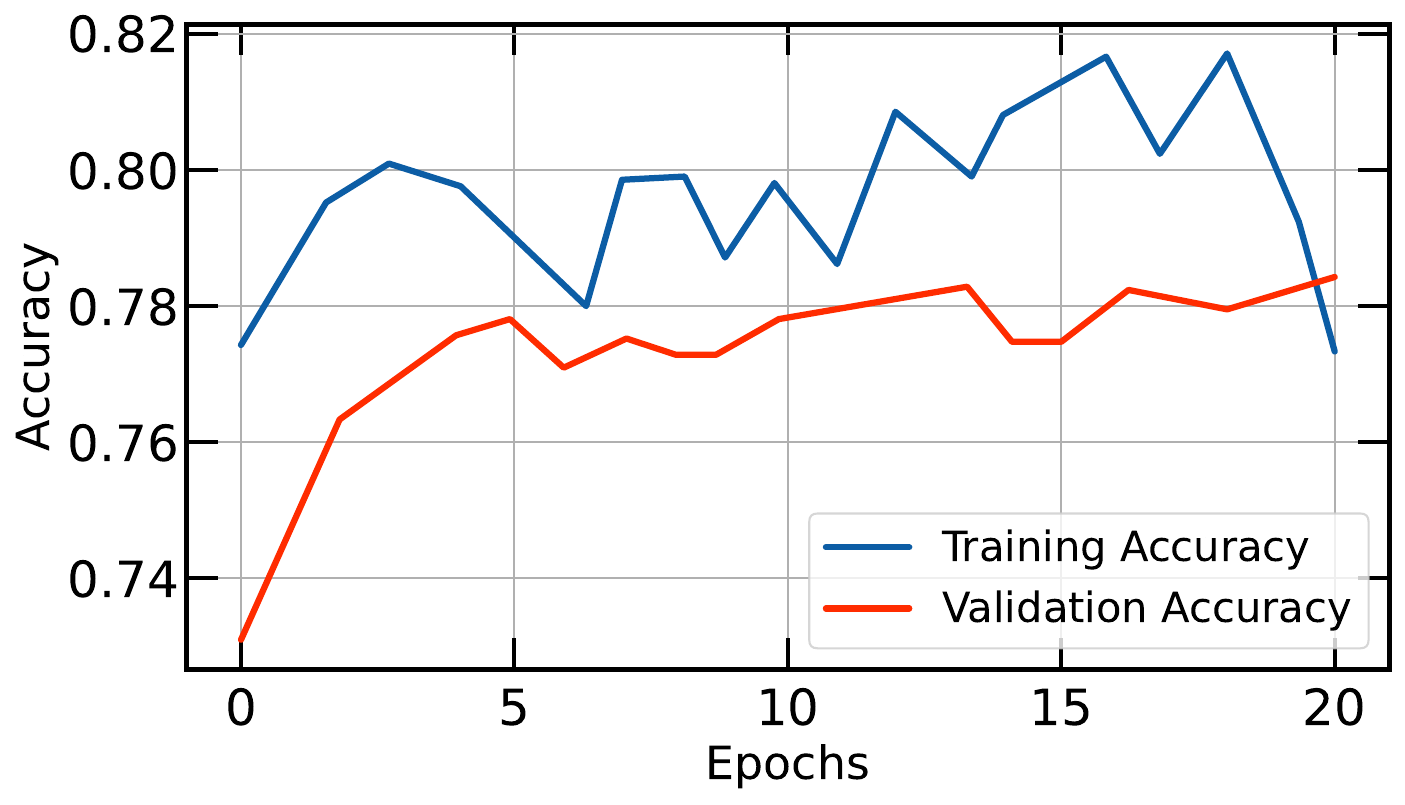}  
    \hfill
    \includegraphics[width=0.2\textwidth]{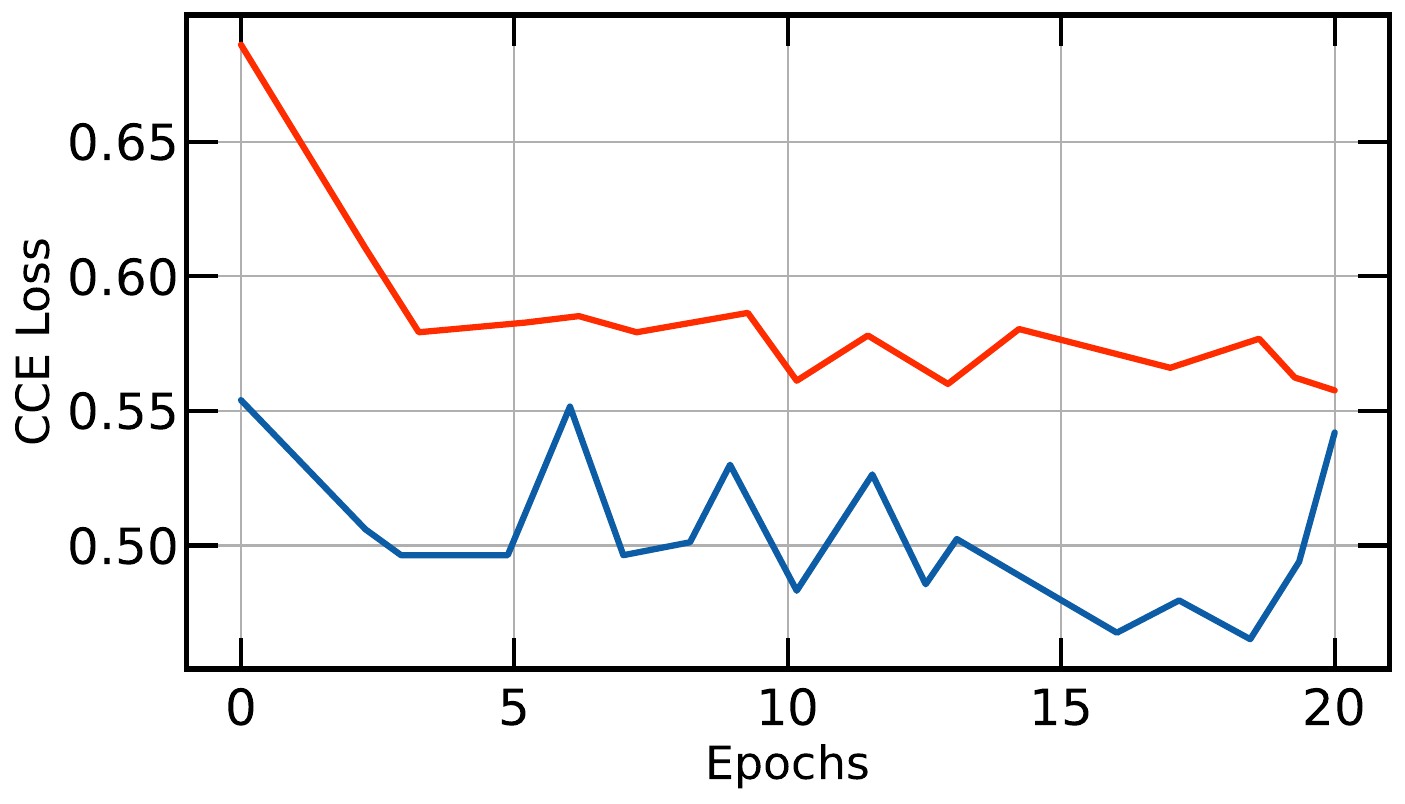}  
    \hfill
    \includegraphics[width=0.2\textwidth]{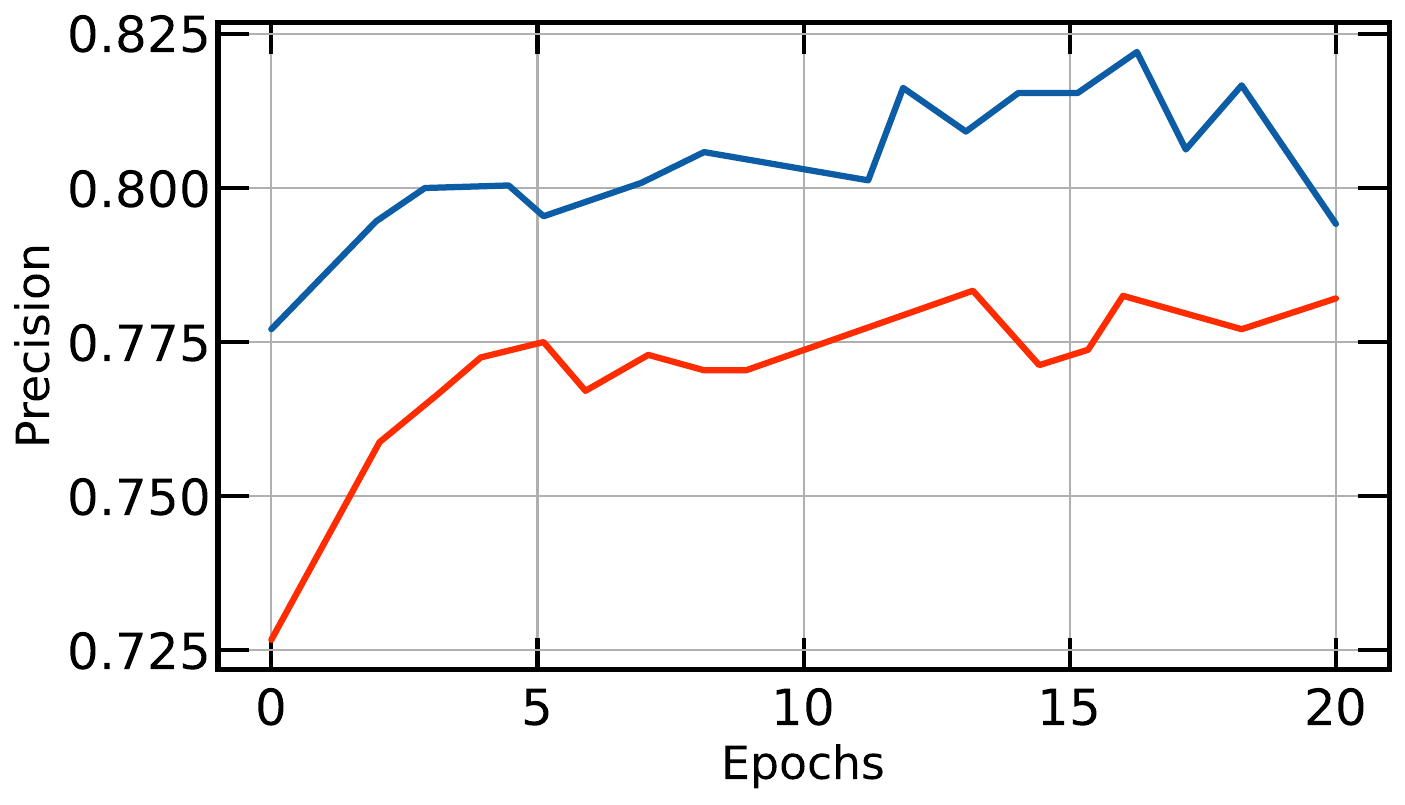}  
    \hfill
    \includegraphics[width=0.2\textwidth]{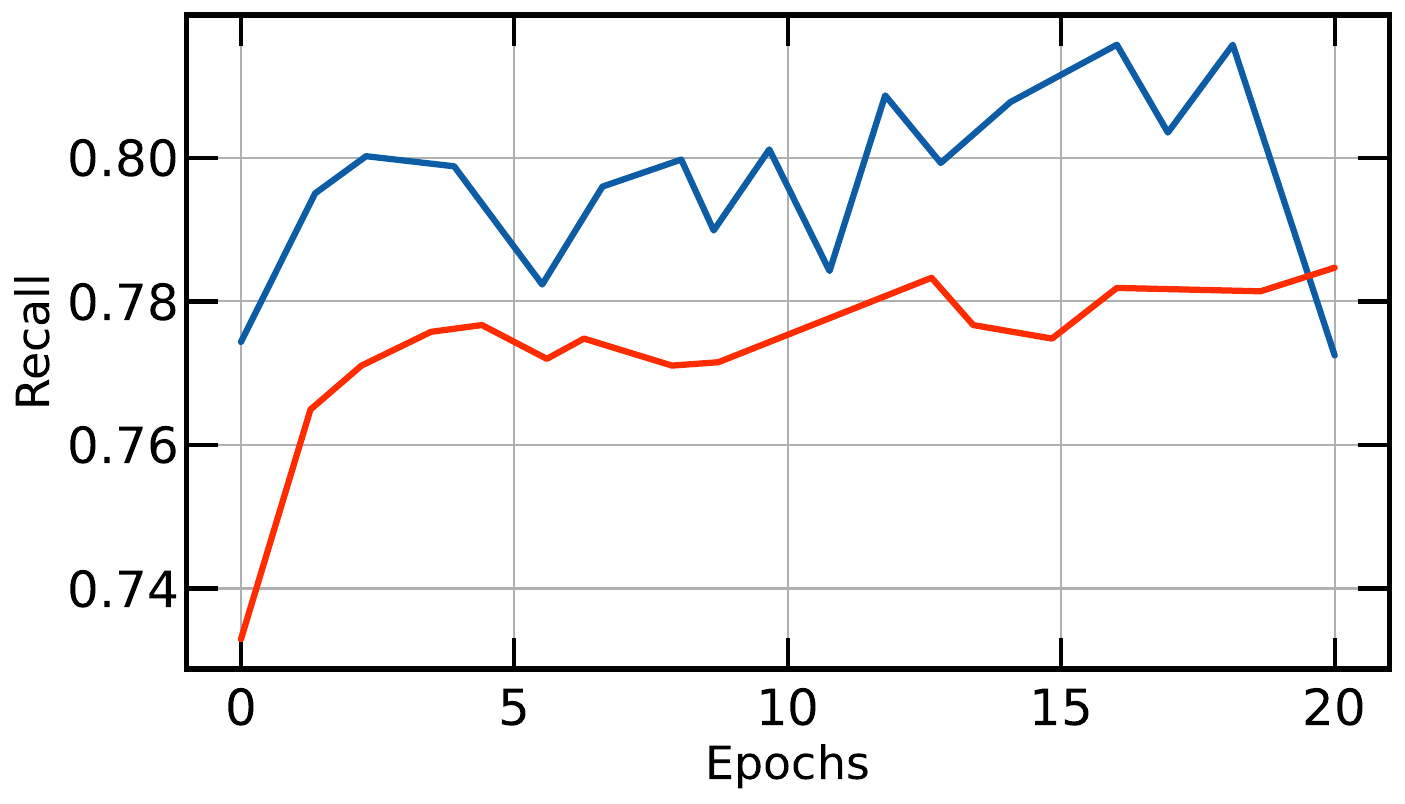}  
    \caption{Performance of VGG19}
    \label{fig: Performance_of_VGG19}
\vspace{-0.3cm}
\end{figure}

\begin{figure}[H]
    \centering
    \includegraphics[width=0.2\textwidth]{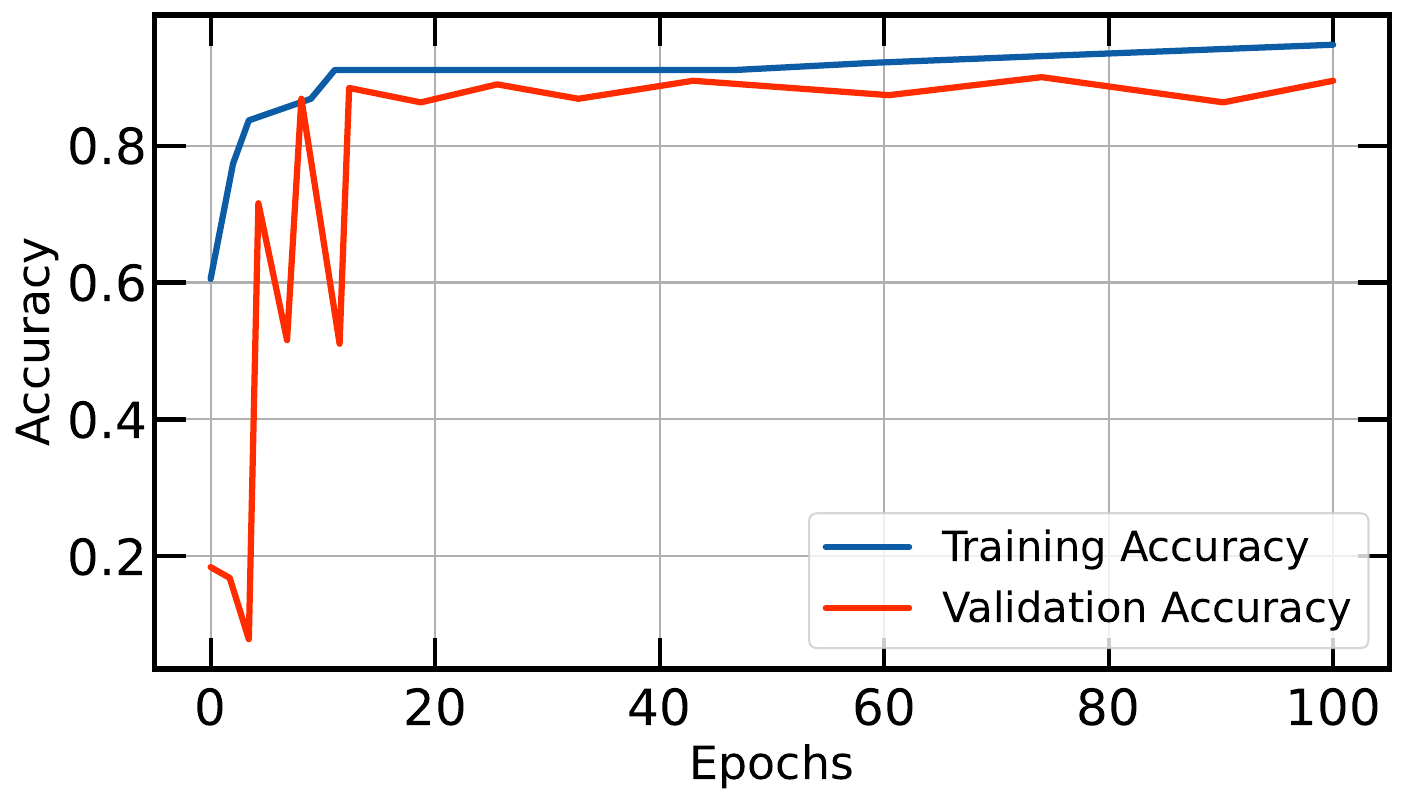}  
    \hfill
    \includegraphics[width=0.2\textwidth]{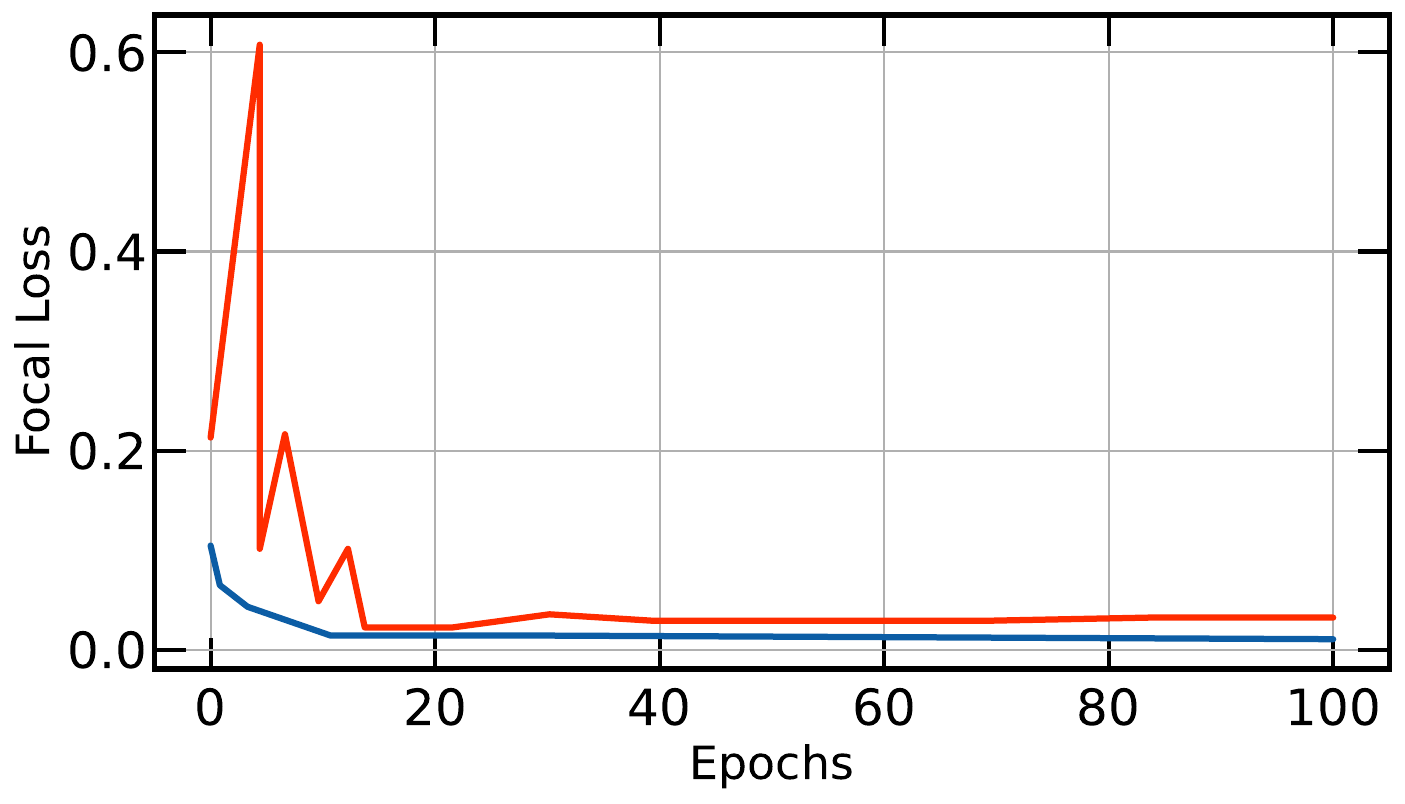}  
    \hfill
    \includegraphics[width=0.2\textwidth]{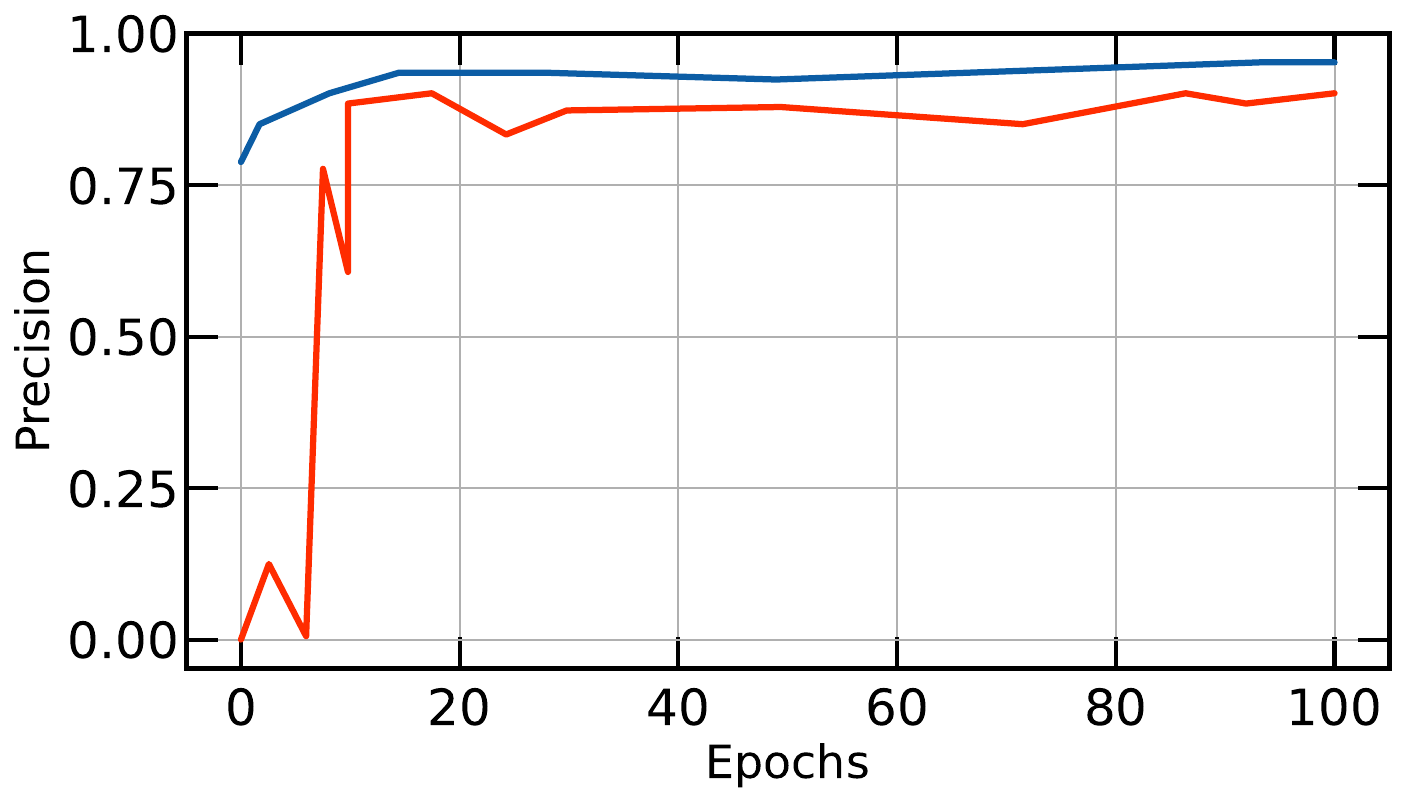}  
    \hfill
    \includegraphics[width=0.2\textwidth]{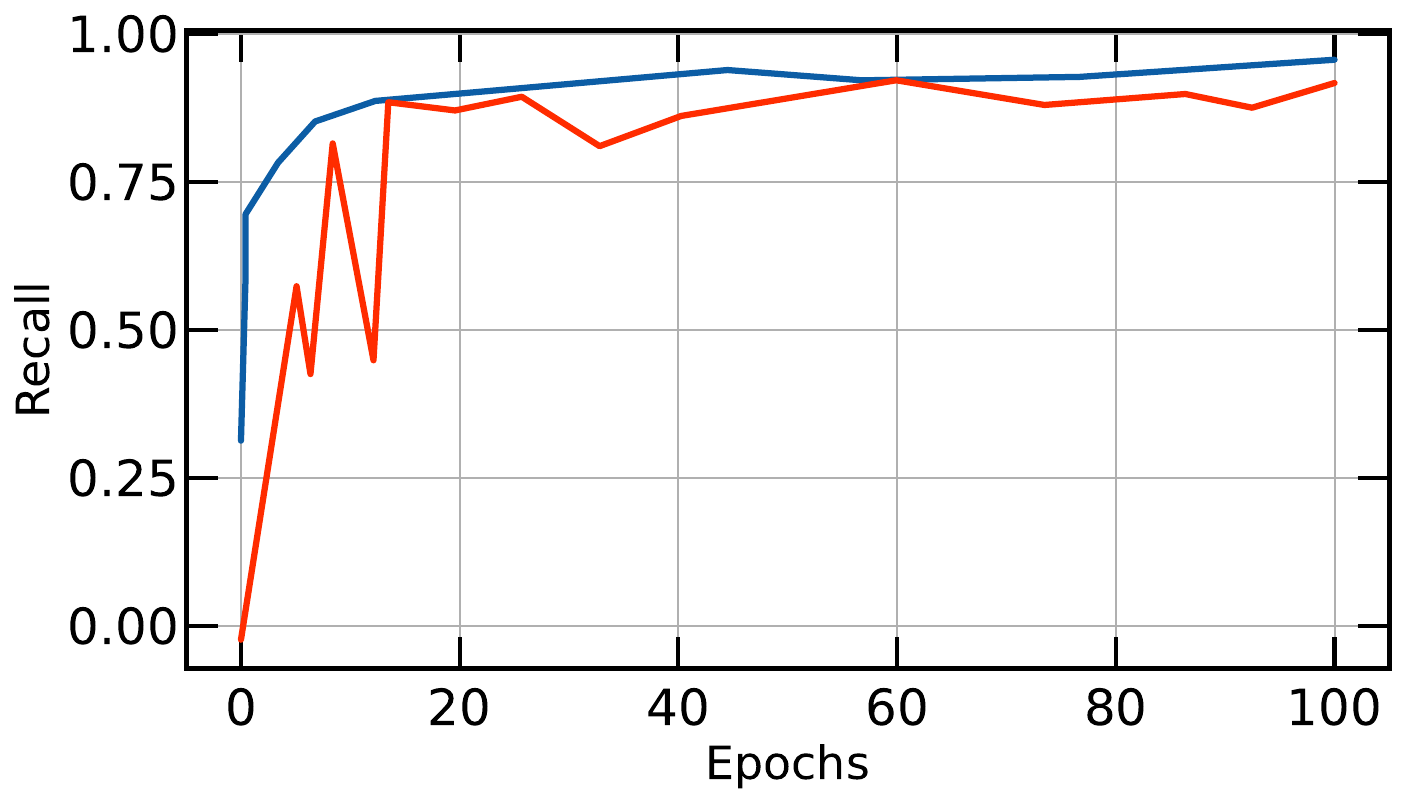}  
    \caption{Performance of Enhanced DenseNet121 with Focal Loss (100 epochs)}
    \label{fig: Performance of Enhanced DenseNet121 with Focal Loss (100 epochs)}
\end{figure}

\section{Conclusions and Future Work}

We have been able to diligently show that adding attention blocks to DenseNet improves its performance significantly in predicting diseases from X-ray images. This owes to the focusing nature of attention block. Furthermore, the depthwise convolution has been shown to decrease the complexity of the mode while preserving its ability to perform classification accurately. Considering that our dataset was not augmented, we deem it a necessary step that will make our model more robust. Furthermore, additional comparisons should be made with other models, especially after adding attention blocks and depthwise convolution.

{
\bibliographystyle{IEEEtran}
\bibliography{ref} 
}
\end{document}